# Unidirectional electron-phonon coupling as a 'fingerprint' of the nematic state in a kagome superconductor

Ping Wu[1#], Yubing Tu[2#], Zhuying Wang[1], Shuikang Yu[1], Hongyu Li[1], Wanru Ma[1], Zuowei Liang[1], Yunmei Zhang[1], Xuechen Zhang[1], Zeyu Li[1], Ye Yang[1], Zhenhua Qiao[1], Jianjun Ying[1,3], Tao Wu[1,3,4], Lei Shan[2], Ziji Xiang[1], Zhenyu Wang[1,3*], and Xianhui Chen[1,3,4*]

[1]Department of Physics, CAS Key Laboratory of Strongly-coupled Quantum Matter Physics, University of Science and Technology of China, Hefei, Anhui 230026, China
[2] Information Materials and Intelligent Sensing Laboratory of Anhui Province, Institutes of Physical Science and Information Technology, Anhui University, Hefei 230601, China
[3] CAS Center for Excellence in Superconducting Electronics (CENSE), Shanghai 200050, China
[4]Collaborative Innovation Center of Advanced Microstructures, Nanjing 210093, China
[#]These authors contributed equally to this work
*Correspondence and requests for materials should be addressed to Z.W. (zywang2@ustc.edu.cn) or X.-H.C. (chenxh@ustc.edu.cn).

**Electronic nematicity has been commonly observed in juxtaposition with unconventional superconductivity. Understanding the nature of the nematic state, as well as its consequence on the electronic band structure and superconductivity, has become a pivotal focus in condensed matter physics. Here we use spectroscopic imaging-scanning tunneling microscopy to visualize how the interacting quasiparticles organize themselves in the nematic state of kagome superconductor $CsV_{3-x}Ti_xSb_5$, in which twofold symmetric ($C_2$) quasiparticle scattering interference of the vanadium kagome bands emerges below the bulk nematic transition temperature ($T_{nem}$). Surprisingly, we find that the coupling to collective modes, i.e., the phonon, dramatically alters the electron's self-energy and renormalizes the Fermi velocity of the in-plane vanadium $d_{xy/x^2-y^2}$ bands only along the $C_2$ direction, making the low-energy dispersion and electron dynamics highly nonequivalent along the three lattice directions. The anti-correlation between $T_{nem}$ and the superconducting transition temperature upon Ti substitution further suggests a possible competition between superconductivity and electron nematicity in this series, with a principal superconducting gap opening on the same V bands once the nematic state is totally suppressed. The organizing principle of these quasiparticles provides essential information for understanding the interplay between charge density wave and superconductivity in these kagome superconductors, and also reveals a previously unexplored form that expands the landscape for modelling electronic nematicity in systems where electron correlations and lattice degree of freedom act in concert.**

Electron liquid in a solid-state system generally respects certain symmetries imposed by the underlying crystal lattice. Many-body interactions, typically electron-boson coupling, would produce a characteristic electron self-energy $\Sigma(\boldsymbol{k}, \omega)$ that encodes the coupling information for electronic states $|\boldsymbol{k}\rangle$ at a given energy $\hbar\omega$. With appreciable interactions, the symmetry of the electron's wave function might be spontaneously broken, resulting in a startling array of electronic phases. An intriguing example is the electronic nematic phase, in which the electrons have a strong tendency to self-organize into a ground state with lower rotational symmetry of the underlying lattice (1-3). Such electronic nematicity is widely observed in systems such as copper- and iron-based high-temperature superconductors (4-12), quantum Hall liquid (13,14), and graphene moiré superlattices (15-17); its driving mechanism and relationship with other symmetry-breaking phases, especially the superconductivity, have become a pivotal focus in recent years. However, a microscopic description of the nematic order, as well as the determination of the dominant interaction responsible for it, is usually difficult to achieve (11). One of the main problems is that the sensitivity of nematic order to structural disorder naturally forms dense spatial domains, so that little information can be obtained from global measurements.

The recently discovered kagome superconductor $A$V$_3$Sb$_5$ ($A$= K, Rb, Cs) brings an excellent opportunity to explore electronic nematicity that intertwines with charge density wave and superconductivity (18-23). This class of materials consists of a two-dimensional vanadium kagome net whose hexagonal centres are filled with Sb atoms (Sb1), whilst additional Sb atoms (Sb2) form honeycomb lattices surrounding the kagome layer (Fig. 1a). There are multiple kagome-derived van Hove singularities located at the $M$ points near the Fermi level, with two types of sublattice flavours (24-26). Upon cooling, the system exhibits a triple-$Q$ $2a_0 \times 2a_0$ charge density wave (CDW) transition at $T_{CDW}$ ~78-104 K, with in-plane wave vectors connecting these van Hove singularities (19,27-32). Indications of time-reversal symmetry breaking have been reported inside the CDW phase without long-range magnetic order (27,32-34), implying the unconventional nature of this triple-$Q$ CDW order. On the other hand, two-fold-symmetric ($C_2$) amplitude of the triple-$Q$ $2a_0 \times 2a_0$ CDW modulation (23, 35, 36) has been found. Strong nematic fluctuations emerge in the CDW phase of CsV$_3$Sb$_5$; they eventually trigger a nematic transition (with respect to the CDW unit-cell) at a lower temperature of $T_{nem}$ ~35 K (23). The $C_2$ rotational symmetry persists into the superconducting state as hinted by transport (37) and STM measurements (28, 36). However, key information on the organizing principle of low-energy quasiparticles in the nematic phase is still lacking. It is also of fundamental interest to examine the nematic phenomena at the microscopic level in this class of kagome compounds where the spin degree of freedom is less essential.

We pursue these objectives by using spectroscopic-imaging scanning tunneling microscopy (SI-STM) to visualize how the quasiparticles organize themselves in the nematic state in Ti-substituted CsV$_3$Sb$_5$. Our high-resolution quasiparticle interference (QPI) measurements detect most of the scattering interference channels that connect the vanadium kagome bands, and the resulting dispersion allows us to determine that a special electron-phonon coupling modifies the quasiparticle dispersion in a highly anisotropic manner with respect to the three lattice directions. The evolution of these quantum interference patterns as a function of both temperature and Ti-content $x$ establishes a direct link between these unidirectional quasiparticles and bulk electronic nematicity, and further suggests that the nematic phase may act as a possible competitor to superconductivity.

We begin with presenting the phase diagram of $CsV_{3-x}Ti_xSb_5$ as determined by bulk transport measurements in Fig. 1b (see Extended Data Fig. 1 for details). The CDW transition can be identified by a kink in the temperature-derivative of the in-plane resistivity; titanium substitution continuously suppresses the CDW order till x $\approx$0.12, above which the CDW transition becomes invisible in $d\rho_{ab}/dT$. Interestingly, the superconducting transition temperature, $T_C$, undergoes a rather complex, non-monotonic evolution with Ti substitution. To gain a deeper insight into the underlying electronic properties, particularly for the evolution of the concealed electronic nematicity, we focus on samples with four representative Ti contents: $x = 0$, $x \approx 0.05$ with the lowest $T_C \sim 0.8K$, $x \approx 0.12$ (above which the CDW order disappears), and $x \approx 0.18$ with the optimized $T_C \sim 3.2K$ (marked with stars in Fig. 1b).

The crystals cleave naturally between the adjacent alkali and Sb layers, exposing either Cs or Sb terminated surfaces (27-30). Figures 1c to j depict typical topographic images of the Sb-terminated surfaces, in which the honeycomb lattice of Sb is visible with lattice constant $a_0$ = 5.5 Å. For the Ti-substituted samples, we find a large number of sharp, atomic-scale defects in the topographs taken at high bias voltages (e.g., $V_S$=1.5 V; Fig. 1c to f); these defects appear as bright protrusions located at half-way between pairs of the topmost Sb atoms, which we identify as the Ti atoms in the underneath kagome layer (see schematic shown in Fig. 1a). By counting the numbers of these defects in large fields of view (typically larger than 50-nm-square; Extended Data Fig. 2), the Ti concentrations can be determined with high accuracy. In fact, the Ti compositions yielded this way are in good agreement with the results of inductively coupled plasma atomic emission spectroscopy (ICP-AES) measurements, and show weak spatial variations in the samples we studied here (Supplementary Table 1), confirming the high quality of these samples.

The topographs taken at low bias voltages (Fig. 1g-j) show the triple-$Q$ $2a_0 \times 2a_0$ CDW order together with a unidirectional $1Q$-$4a_0$ charge modulation (sometimes $5a_0$ due to phase slips, ref. 28-30) at 4.8K. These two types of charge order can be clearly observed as peaks at the corresponding wave vectors in the Fourier transforms (FTs) of the topographs, as shown in a typical example in Fig. 1k. To illustrate the evolution of these charge orders with Ti substitution, we present the FT linecuts along the $1Q$-ordering direction for samples with different $x$ in Figure 1l (normalized to the amplitude of associated Bragg peaks). The $1Q$-$4a_0$ charge order, observed only on the Sb surfaces, tends to locally melt with Ti-substitution (Fig. 1h), and eventually disappears in the $x \approx 0.12$ samples (Fig. 1i and l); the triple-$Q$ $2a_0 \times 2a_0$ CDW order, on the other hand, is still distinguishable for $x \approx 0.12$ and vanishes at larger $x$, consistent with the transport measurements.

To reveal the electronic band structures, we next implement the technique of QPI imaging to CsV$_{3-x}$Ti$_x$Sb$_5$. When quasiparticles in crystals are scattered by impurities/crystal defects, they quantum-mechanically interfere to produce spatial modulations in the density of states $\delta N(\bm{r}, \omega)$ at characteristic scattering vectors $\bm{q}$. Each $\bm{q}$-vector that connects two electronic states $|\bm{k}\rangle$ and $|\bm{k}+\bm{q}\rangle$ in the momentum-space can be extracted from the Fourier-transformed image, $\delta N(\bm{q}, \omega)$. Technically, these interference patterns $\delta N(\bm{r}, \omega)$ are measured by differential conductance maps $dI/dV(\bm{r}, eV) \equiv g(\bm{r}, eV)$ as a function of location $\bm{r}$ and bias voltage $V$, where $I$ is current and $e$ is the electron charge. Here, we typically measure g($\bm{r}, eV$) in large fields of view between 60-nm-square and 100-nm-square, to achieve a high $\bm{q}$-space resolution of $|\delta q| \sim 0.007(2\pi/a_0)$; we also use normalized conductance maps in the Feenstra form, $L(\bm{r}, eV) \equiv g(\bm{r}, eV)/(I(\bm{r}, V)/V)$, to suppress the systematic error related to the set-point effect in the constant-current scanning mode (38, 39; see Method). The high-precision imaging of QPI can not only be used to measure details of the electronic band dispersion $\bm{k}_i(\omega)$, but also provide a rigorous test for both the quasiparticles' coherence (40) and the underlying symmetry-breaking (7, 9, 39).

Figures 2, a to c, show typical examples of atomically registered L($\bm{r}, eV$) maps for samples with $x = 0$, $x \approx 0.05$, and $x \approx 0.18$, respectively, at 4.8 K above the superconducting transitions. Fourier transforms (FTs) of the maps, L($\bm{q}, eV$), display a rich array of scattering events as shown in Fig. 2 d-f, where a sequence of salient scattering channels are labelled as $\bm{q}_i$: i=0,1,2,3,4,5. To identify the electronic band elements responsible for these interferences, we consider the band structure of CsV$_3$Sb$_5$ as outlined in Fig. 2g, which has been established by *ab-initio* calculations and angle-resolved photoemission (ARPES) measurements (19, 24, 41). The constant energy contours consist of a circular Sb-$p$ pocket (α) around $\Gamma$, a large hexagonal hole-like pocket (β) centered at $\Gamma$ with V-$d_{xz}/d_{yz}$ orbital character, a triangular V-$d_{xz}/d_{yz}$ pocket (γ) and a triangular electron-like V-$d_{xy}/d_{x^2-y^2}$ pocket (δ) surrounding the $K$ point. Here we mainly focus on the filled states, in an energy window where the β and γ bands are very close to each other (Fig. 2g; -40 meV) and difficult to distinguish with QPI measurements. To achieve a clearer QPI prediction, we next consider a joint density-of-states (JDOS) approximation for the quasiparticles of vanadium bands, and show the resulting interference patterns in Fig. 2h. Careful examination of the directions and lengths of the observed $\bm{q}_i$ allows us to identify their origins: $\bm{q}_0$ represents the intra-band scattering of the Sb-$p$ pocket; $\bm{q}_1$ ($\bm{q}_2$) represents the scattering between the parallel edges (tips) of triangular δ pockets; $\bm{q}_3$ originates from the intra-β (γ)-band scattering; $\bm{q}_4$ ($\bm{q}_5$) is equivalent to $\bm{q}_1$ ($\bm{q}_3$) up to a reciprocal lattice vector. We note here that while $\bm{q}_1$ and $\bm{q}_3$ share similar $\bm{q}$-space regions near the Fermi energy (E$_F$), we can still track them in the high-resolution $L(\bm{q}, eV)$ maps by their different curvatures resulting from the shape of the bands (i.e., see Fig. 2d1 and Supplementary Note 1). One key observation here is that the interference patterns associated with vanadium $d$-bands are highly unidirectional, for samples of $x = 0$ and $x \approx 0.05$, with striking $C_2$-symmetric scattering signals along a single lattice direction (labeled as $C_2$-direction hereafter); for $x \approx 0.18$, however, the sixfold-rotational symmetry is restored (raw data shown in Fig. 2c and f1). As we will discuss later, this phenomenon is intimately related to the electronic nematicity. The L(*r*, *eV*) maps showing dense spatial $C_2$ domains on the x~0.12 samples and their FTs are displayed in Supplementary Note 2.

A primary advantage of the present L(***q***, *eV*) datasets is that the scattering channels ***q***$_i$ can be tracked to high energies (up to -100 meV in $x \approx 0.05$ samples; complete sets of data are shown in Supplementary Movies 1-4). The electronic band structures $\mathbf{k}_i(\omega)$, can then be determined by the measured ***q***$_i$ geometrically through $2|\mathbf{k}_1^\delta(\omega)| + |\mathbf{q}_1(\omega)| = \sqrt{3}\,(2\pi/a_0)$, $2|\mathbf{k}_3^\beta(\omega)| = |\mathbf{q}_3(\omega)|$ along $\Gamma K$, and $2|\mathbf{k}_2^\delta(\omega)| = |\mathbf{q}_2(\omega)|$ along the $\Gamma M$ direction. The extracted $\mathbf{k}_i(\omega)$ are presented in Fig. 3a, together with the calculated band dispersion considering a $2a_0 \times 2a_0 \times 2c$ CDW reconstruction (at $k_z = \pi/2c$). In fact, a three-dimensional CDW order, with tri-hexagonal reconstruction in the kagome plane and a π-phase shift between the neighboring planes, has been suggested in CsV$_3$Sb$_5$ by XRD (42), nuclear magnetic resonance (23) and nuclear quadrupole resonance measurements (43, 44). As such, we consider this type of CDW configuration and use the experimental lattice parameters determined by XRD at 18 K from Ref. 42 in the calculation. One can find excellent agreement between the data and calculations, which further bears out the validity of the identifications of ***q***$_i$ discussed in Fig. 2. More calculations with different CDW configurations are shown in Extended Data Fig. 3.

Having understood the dominant ***q***-vectors, we now turn to the unidirectionality of the interference signals raised from the vanadium kagome bands. In a more rigorous description employing T-matrix formalism, the interference pattern δN (***q***, ω) can be predicted as

$$\delta N(\mathbf{q}, \omega) = -\pi^{-1}\mathrm{Im} \int d\mathbf{k}\, G_s(\mathbf{k}, \omega) T(\mathbf{k}, \mathbf{k}+\mathbf{q}, \omega) G_s(\mathbf{k}+\mathbf{q}, \omega)$$

with $G_s(\mathbf{k}, \omega)$ the electron Green's function in orbital space and $T(\mathbf{k}, \mathbf{k}+\mathbf{q}, \omega)$ the matrix representing all the possible scattering channels between $|\mathbf{k}\rangle$ and $|\mathbf{k}+\mathbf{q}\rangle$. Both an anisotropic T-matrix and a 'self-energy' correction to the electron Green's function can lead to an anisotropic δN (***q***, ω), as shown in the framework of iron pnictides (9, 45). To gain insight into these two possibilities, we show, in Fig. 3b and c, the L(***q***, ω = -60 meV) map obtained in the $x \approx 0.05$ samples, and compare the magnitudes of ***q***-vectors with respect to the $C_2$-direction (pink block) and one of the remaining two (off-$C_2$-direction; dark blue block) in Fig. 3c. Interestingly, while the lengths of the reciprocal lattice vector, the 2a$_0$ × 2a$_0$ CDW vector, and the intra-α-band (***q***$_0$; Sb-*p* orbital) scattering vector are all identical (within one-pixel deviation), the scattering vectors related to the vanadium *d*-band (***q***$_1$ and ***q***$_3$) show clear differences, indicating non-equivalent electronic states $|\mathbf{k}\rangle$ along these two directions. This finding directs our focus to the energy dispersion of $k_1^\delta(\omega)$, which has primarily in-plane $d_{xy}/d_{x^2-y^2}$ orbital character and shows the largest anisotropy in the data (the fitting procedure using Lorentzian line-shape to extract $k_1^\delta(\omega)$ is shown in Supplementary Note 3). The results, shown in Fig. 3d and e, reveal a key observation: while the measured dispersion of $k_1^\delta(\omega)$ along the off-$C_2$-direction roughly follows the calculated "bare" band, the one with respect to the $C_2$-direction shows visible kinks at ~ 15 and 30-40 meV in the dispersion associated with a reduced Fermi velocity----a hallmark of electronic 'self-energy' $\Sigma(k, \omega)$ renormalizations (the kink at ~15meV is more visible in the zoom-in shown in Supplementary Note 3). The real part $\Sigma'(k, \omega)$, which describes the changes in the dispersion, is then plotted in Extended Data Fig. 4 by subtracting the bare band, showing clear peaks located at approximately 15 and 30-40 meV along the $C_2$-direction but almost negligible effects along the other two directions. We emphasize that this observation is not a measurement artifact, but has been confirmed in five areas across three $x \approx 0.05$ samples with different orientations of the scanning frame (Fig. 3e). A similar phenomenon is also found in the parent $x = 0$ samples (Extended Data Fig. 5). For further confirmation of the $C_2$-symmetric band renormalization, it is necessary to check the energy-dependent peak widths and peak heights of $k_1^\delta(\omega)$, which are in principle related to the imaginary part of the quasiparticle self-energy $\Sigma''(k, \omega)$. The quasiparticle linewidths, obtained from the phenomenological full-width-half-maximum of the QPI signal ($\Delta k = \Delta q/2$), are plotted in Extended Data Fig. 6. While the energy-dependent linewidths are rather featureless along the off-$C_2$-direction, those along the $C_2$-direction clearly show a more rapid decrease below 40 meV, which corresponds to the kink position in the dispersion. Furthermore, the difference in peak heights along these two directions also exhibits a significant enhancement at energies $|eV| < 40$ meV (Extended Data Fig. 6). This again is consistent with the case of $C_2$-symmetric electron-mode coupling, complementing the dispersion data in Fig. 3. Therefore, the extracted dispersion $k_1^\delta(\omega)$, the energy-dependent linewidths and peak heights taken together unambiguously demonstrate a unidirectional self-energy renormalization in lightly (none)-doped $CsV_{3-x}Ti_xSb_5$ at low temperature.

One might wonder whether the unidirectional band structure is a simple consequence of the 1$Q$-4$a_0$ charge order observed on the Sb surface, unrelated to the bulk electronic nematicity. To clarify this, we performed temperature-dependent measurements. For the $x = 0$ sample, we find that while the quantum interference pattern related to the Sb-$p$ orbital ($q_0$) persists to high temperature, those derived from the vanadium band vanish above approximately 40 K (Fig. 4a and c), a temperature close to $T_{nem}$ ~35 K (ref. 23) but much lower than the onset temperature of the 4$a_0$ order (~ 55K; Extended Data Fig. 7), consistent with previous reports (28, 36). On the other hand, while the onset temperature of the 4$a_0$ order is suppressed to below 40 K for $x \approx 0.05$ (one can also see the local melt of this order in Fig. 1h), the emergent temperature of the $C_2$ QPI patterns increases to above 50 K (Fig. 4b and d), consistent with an increase in $T_{nem}$ revealed by elastoresistance measurements on samples with similar doping. The opposite evolution of these two temperature scales provides strong evidence that the unidirectional QPI signals are intimately related to the electronic nematicity in which a $C_2$-symmetric ground state emerges. Further support comes from a comparison of the self-energy renormalization between $x \approx 0.05$ and $x = 0$: although the Ti substitution suppresses the $1Q$-$4a_0$ order for $x \approx 0.05$, the extracted $\Sigma'(\mathbf{k}, \omega)$ is more remarkable compared to that of the pristine samples (Extended Data Fig. 4).

The unidirectional self-energy correction results in an enhancement of quasiparticles' effective mass $m^*$ by a factor of ~2 along the $C_2$-direction on the vanadium $d_{xy}/d_{x^2-y^2}$ band ($v_0^\delta \sim 3.1 \times 10^5$ $m/s$ from calculations and $v_{C2}^\delta \sim 1.6$ ($\pm 0.3$) $\times 10^5$ $m/s$ from data). This leaves us with the question of what type of interaction could be responsible for it. To proceed, we first check the calculated phonon DOS of $CsV_3Sb_5$ (Extended Data Fig. 8), which predominantly shows a series of peaks in similar energy windows where we see kinks: some near 15 meV and others between 30 and 40 meV. The amplitudes of these peaks are largely enhanced when entering the CDW phase (calculated for the tri-hexagonal configuration). Experimentally, neutron scattering finds a weakly dispersive $B_{3u}$ phonon mode (half-breathing mode of vanadium) at $E \approx 10$ meV that hardens across $T_{CDW}$ at the $M$ point (46). Several phonon modes at frequencies of 13 and 30 meV have been observed below approximately 40 K by Raman spectroscopy and time-resolved spectroscopy (31, 47-49). These observations suggest phonon as the prime collective mode.

This immediately raises the question of how to understand the emergence temperature for the $C_2$ interference pattern of the vanadium bands and the unidirectional electron-phonon coupling. In general, a triple-$Q$ CDW order can be described by a multi-component order parameter ($\psi_1$, $\psi_2$, $\psi_3$), with $\psi_i$ ($i = 1, 2, 3$) representing the CDW component along the three lattice directions (50, 51). The CDW order preserves the rotational symmetry of kagome lattice when $\psi_1 = \psi_2 = \psi_3$. There is accumulated evidence showing that the rotational symmetry breaking is explicitly broken at $T_{CDW}$, owing to the $C_2$ structural distortion with the formation of three-dimensional CDW (23, 33, 36). This $C_2$ structural distortion with $\psi_1 = \psi_2 \cong \psi_3$, is not associated with detectable electronic nematicity just below $T_{CDW}$, as suggested by nuclear spin-lattice relaxation rate and STM measurements (23). On the other hand, the anisotropy of electronic states gets largely enhanced when electronic nematicity with remarkable $\psi_1 = \psi_2 \neq \psi_3$ emerges below a lower temperature $T_{nem}$, without any resolvable structural transition (23, 42); nematic fluctuations have been observed between $T_{CDW}$ and $T_{nem}$. The emergence of interference patterns of the vanadium bands below $T_{nem}$ strongly suggests the formation of a coherent $C_2$-symmetric electronic state in response to the nematic ordering; consistently, the disappearance of interference patterns above $T_{nem}$ is probably due to the strong nematic fluctuations. Theoretically, the CDW order could be more specifically a charge bond order in which charge modulations occur on the atomic bonds (51-53) when considering the inherent scattering of the kagome lattice at van Hove filling, and the interplay among $\psi_i$ ($i = 1, 2, 3$) would lead to a CDW state that is prone to nematicity or time-reversal symmetry breaking (51-53). In this context, the electronic nematicity below ~35K could be an indicator of the change of the triple-$Q$ CDW order at low temperature that stems from the delicate cooperation of electron-phonon interaction, electron-electron interaction and inherent geometrical frustration of the kagome lattice. The anisotropic bond correlations, as well as the enhanced electronic density of states associated with the $d_{xy}/d_{x^2-y^2}$ higher-order van Hove singularity (24), could generate a striking unidirectional electron-phonon coupling as the fingerprint of this nematic state. Interestingly, recent transport measurements reveal that an electronic magnetochiral anisotropy signal becomes significant only below T*~35 K (54) in the centrosymmetric CsV$_3$Sb$_5$, providing further support to the transition within the CDW phase at this temperature and a strong coupling of the low-temperature CDW phase to the itinerant quasiparticles. This understanding also gives a possible explanation for the higher emergence temperature of $C_2$ CDW on the Sb surface (36) compared with the Cs surface (23), since Sb atoms are directly bonded to the vanadium kagome lattice and thus more sensitive to the structural distortion at $T_{CDW}$. Finally, we would like to note that a $q$=0 Pomeranchuk instability may also play a certain role in the nematic electronic phase (25); however, its effect is difficult to systematically address in our data (Extended Data Fig. 5 and Fig. 3e) and needs further study.

Our results in Figs.2-4 reveal that the electronic nematicity in CsV$_3$Sb$_5$ manifests as an electron-coherent state emerging deep within the CDW phase, with unidirectional self-energy corrections on the kagome-derived bands. The electron-mode coupling strength $\lambda$, estimated from the ratio of the Fermi velocities of the bare band and the dressed quasiparticles ($\lambda = v_0^\delta/v_{data}^\delta - 1$), reaches as high as ~1 along the $C_2$-lattice direction, a similar value to that observed in many cuprate superconductors (55); however, along the other two lattice directions, the self-energy effect is nearly negligible within our experimental resolution. The unidirectional electron-mode coupling observed here is in contrast to the anisotropic electron-phonon interactions observed in cuprates, in which the difference is manifested between symmetrically inequivalent directions (56), or in graphite intercalation compounds, where it occurs in momentum-space with totally different nesting conditions (57).

Because superconductivity develops in this nematic electronic phase, it is crucial to elucidate the relationship between these two orders. Returning to the global phase diagram of CsV$_{3-x}$Ti$_x$Sb$_5$ (Fig. 1b) and the temperature-dependence of the unidirectional QPI signal (Fig. 4a-d), it seems plausible that the nematic order acts as a competitor to the superconductivity, say, an increase in $T_{nem}$ would correspond to a suppression of the superconducting transition temperature. In this case, when the nematic order vanishes with doping, one would expect to see a clear superconducting gap opening on the vanadium bands. The *dI/dV* spectrum of $x \approx$0.18 (with sixfold-symmetric normal-state QPI) shows a larger superconducting gap with a "U"-shaped bottom at the Fermi energy, indicating a rather isotropic superconducting gap (58). To further study the momentum-dependence of Cooper-pair formation, we perform Bogoliubov quasiparticle scattering interference (BQPI) imaging on the $x \approx$0.18 samples under perpendicular magnetic fields of 0 and 0.04 T (Extended Data Fig. 9).

Here the application of a small magnetic field (0.04 T) is to excite more Bogoliubov quasiparticles for interference scattering. Several BQPI images obtained at 0.04 T are shown in Fig. 4f. No new scattering channel is observed at energies within the superconducting gap. Visual inspection finds more remaining spectral weights on the Sb-$p$ band deeply inside the gap (e.g., see g($q$, 0 meV)). To confirm this, we carry out quantitative analysis to track the energy-dependent spectral weight g($q_i$, $eV$) within the relevant $q$-space ranges around $q_0$, $q_1$ and $q_2$ (see Method for details), and find a faster suppression of the spectral weight on the vanadium bands ($q_1$ and $q_2$) than on the Sb-$p$ band ($q_0$) inside the superconducting gap, for both the cases of 0 T and 0.04 T (Extended Data Fig. 9). These observations suggest that an isotropic superconducting gap opens on the vanadium bands for $x \approx 0.18$, with gap magnitude at least similar to that on the Sb band. In contrast, the BQPI images of $x = 0$ show clear spectral weight at the vector of $q_2$ even at 0 meV (Extended Data Fig. 10), suggesting one gap minimum located on the vanadium band around the $M$ point.

Overall, the data presented here report an interesting and previously unexplored form of electronic nematicity in a nonmagnetic kagome system near the van Hove filling, and also adds a new piece of information for understanding the interplay among CDW, superconductivity and nematicity with unidirectional electron-mode coupling. Future experiments on detwinned samples involving more techniques, as well as theoretical modelling, are necessary to unravel more details of the nematic state in this kagome superconductor and to see how a unidirectional self-energy correction could contribute to creating novel quantum states.

# Figure 1

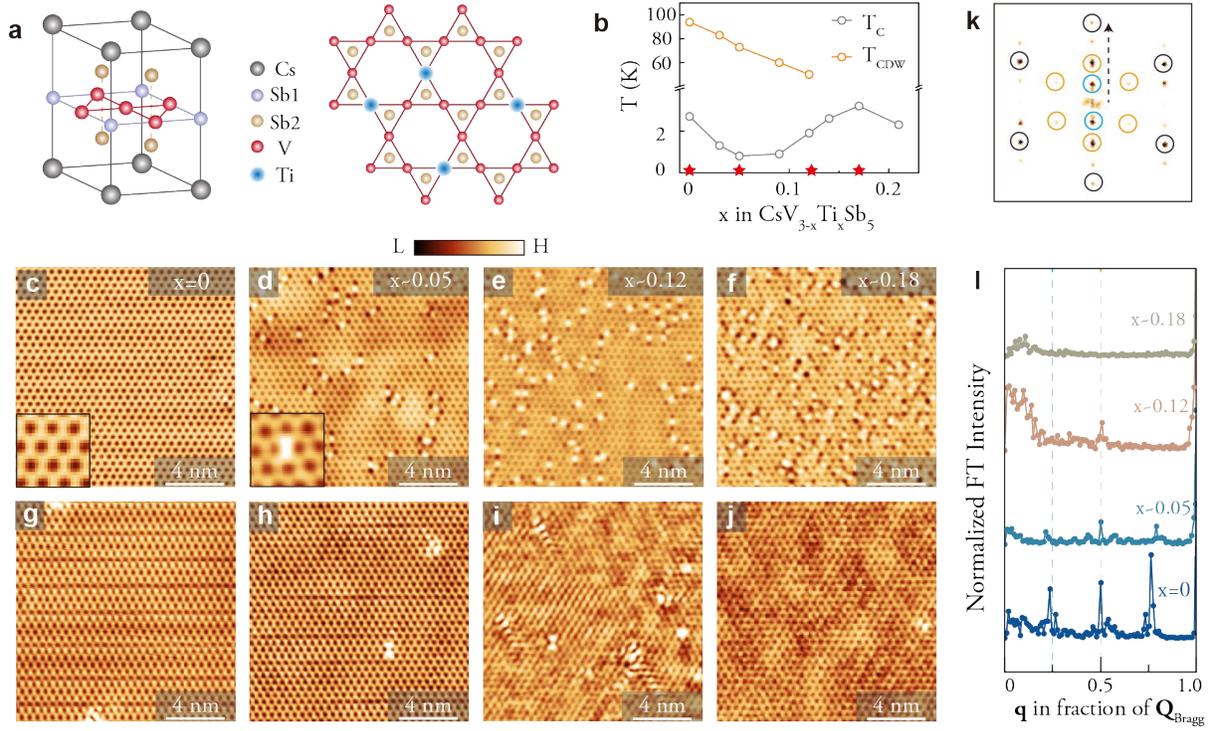

**Figure 1 Phase diagram of Ti-substituted CsV₃Sb₅ and the evolution of charge orders. a,** Crystal structure of CsV₃Sb₅ and a schematic showing the position of substituted Ti atoms in the V₃Sb slabs. **b,** Phase diagram of CsV$_{3-x}$Ti$_x$Sb$_5$ as determined by bulk transport measurements. The stars mark the Ti-contents of the samples on which we performed the STM study. **c-f,** Typical topographs on the Sb-terminated surfaces. The insets in **c** and **d** show the Sb-honeycomb lattice and a Ti dopant located at half-way between pairs of the topmost Sb atoms, respectively. **g-j,** Topographs taken at low bias voltages to show the possible charge modulations, **k,** Fourier transform (FT) of an STM topography of the parent compound. The lattice peaks are marked with black circles, the triple-$Q$ $2a_0 \times 2a_0$ CDW peaks with orange circles, and the $1Q$-$4a_0$ charge order with blue circles. **l,** FT linecuts of STM topographic images along the $1Q$-$4a_0$ ordering direction (if it exists), for samples with four different Ti contents. STM setup condition: **c:** $V_s = 1.5$ V, $I_t = 6$ nA, **d:** $V_s = 1.5$ V, $I_t = 4$ nA, **e:** $V_s = 1.5$ V, $I_t = 2$ nA, **f:** $V_s = 1.5$ V, $I_t = 5$ nA, **g:** $V_s = -0.3$ V, $I_t = 5$ nA, **h:** $V_s = -0.3$ V, $I_t = 4$ nA, **i:** $V_s = -50$ mV, $I_t = 500$ pA, **j:** $V_s = -70$ mV, $I_t = 3$ nA.

# Figure 2

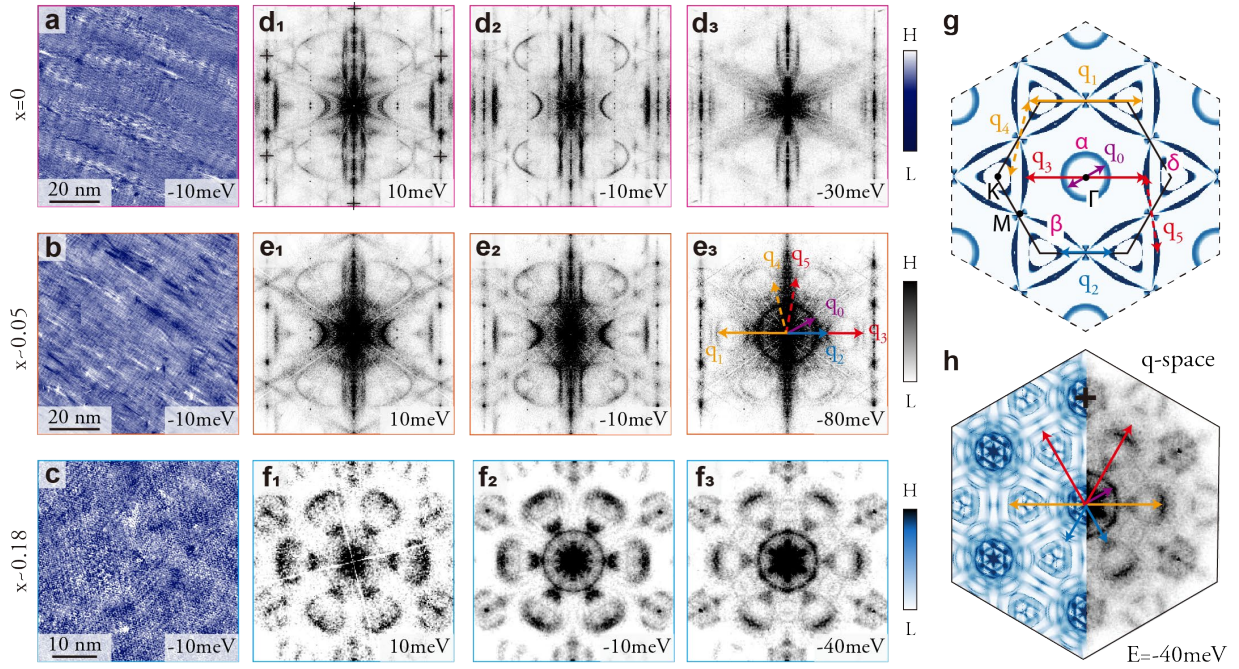

**Figure 2. Quasiparticle interference of CsV$_{3-x}$Ti$_x$Sb$_5$. a-c,** Typical differential conductance $L(r, -10\text{meV})$ maps for the x = 0, x ≈ 0.05 and x ≈ 0.18 samples. **d-f,** Fourier transforms of differential conductance maps, $L(q, eV)$, at several representative energies. Characteristic scattering wave vectors $q_i$: i= 0,1,2,3,4,5 are shown as colored arrows (same arrows in **g** and **h**). The map shown in **f1** is the raw data and the others are two-fold (**d** and **e**) or six-fold (**f2** and **f3**) symmetrized to increase the signal-noise ratio. **g,** Calculated constant energy contours of CsV$_3$Sb$_5$ at -40 meV. Centered at $\Gamma$ are a circular Sb-$p$ pocket (α) and a large hexagonal hole-like pocket (β) with V-$d_{xz}/d_{yz}$ orbital character; surrounding the $K$ point is a triangular electron-like V-$d_{xy}/d_{x^2-y^2}$ pocket (δ). The dark hexagon denotes the first Brillouin zone. For a better view, the constant energy contours are shown in an extended zone. **h,** Predicted QPI pattern using the JDOS approximation (left) in comparison with the measured $L(q, eV)$ (same as in **f3**). **q$_0$** represents the scattering of the Sb-p pocket; **q$_1$ (q$_2$)** represents the scattering between the parallel edges (tips) of triangular δ pockets; **q$_3$** originates from the intra-β-band scattering; **q$_4$ (q$_5$)** is equivalent to q$_1$ (q$_3$) up to a reciprocal lattice vector. The crosses in **d1** and **h** denote the reciprocal lattice vector (Bragg peaks).

# Figure 3

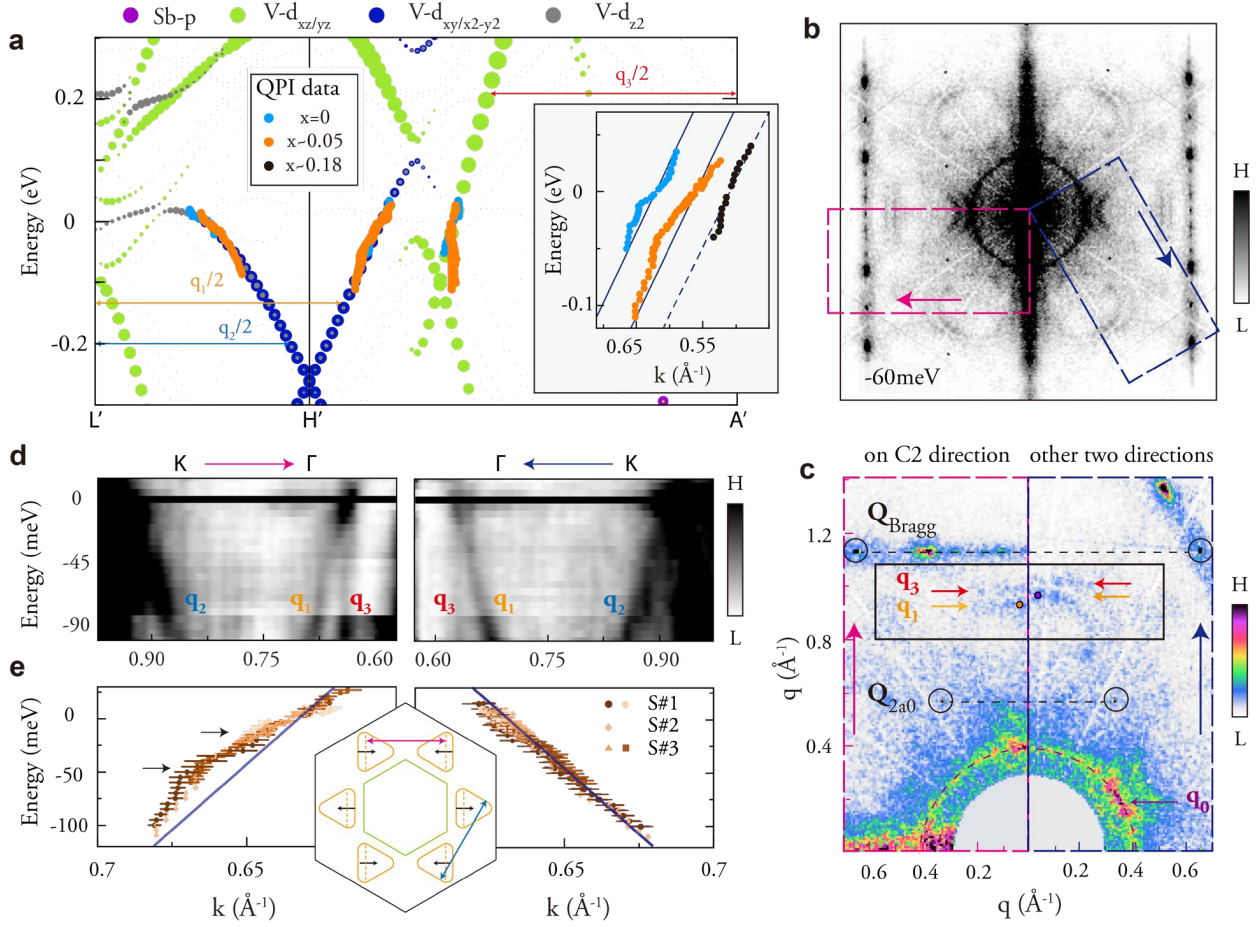

**Figure 3. Visualizing the unidirectional electron-mode coupling on the V-$d_{xy}/d_{x^2-y^2}$ bands.** **a**, Measured $k_i(\omega)$ for the vanadium bands and the calculated band dispersion at $k_z = \pi/2c$, with 2 ×2 ×2 CDW reconstruction (Method). The dispersion is calculated using the experimental lattice parameters determined in Ref. 42. The inset: zoom-in of $k_1^\delta(\omega)$ for x = 0, x ≈ 0.05 and x ≈ 0.18 (horizontally shift for clarification). **b**, The measured $L(q, -60\text{ meV})$ map for x ≈ 0.05. The pink and dark blue rectangles mark the $q$-space regime for comparison. **c**, A direct comparison of the interference patterns with respect to the $C_2$-direction (pink) and that of the other directions (dark blue). The magnitudes of scattering vectors related to the vanadium bands show clear differences ($q_1$ and $q_3$) along these two directions. **d**, Linecuts of $L(q, \text{eV})$ maps as function of energy to track the dispersion of $k_1^\delta(\omega)$ for x ≈ 0.05, along $\Gamma K$ with respect to the $C_2$-direction (left) and off-$C_2$ direction (right). **e**, Measured $k_1^\delta(\omega)$ dispersion in five areas across three different x ≈ 0.05 samples showing excellent consistency. Kinks at approximately 15 and 30-40 meV are observed only along the $C_2$-direction (marked by dark arrows; these kinks are more visible in the zoom-in (SI note 3) and in the self-energy plots (Extended Data Fig. 4)); along the remaining two lattice directions, the measured $k_1^\delta(\omega)$ roughly follows the 'bare' band with no additional feature. The error bars are the sum of the standard deviation widths from fitting and the pixel resolution. The inset shows a schematic representation of the band structure anisotropy at kink energies. The pink (blue) arrow denotes the scattering vector of $q_1$ on (off) the $C_2$-direction.

# Figure 4

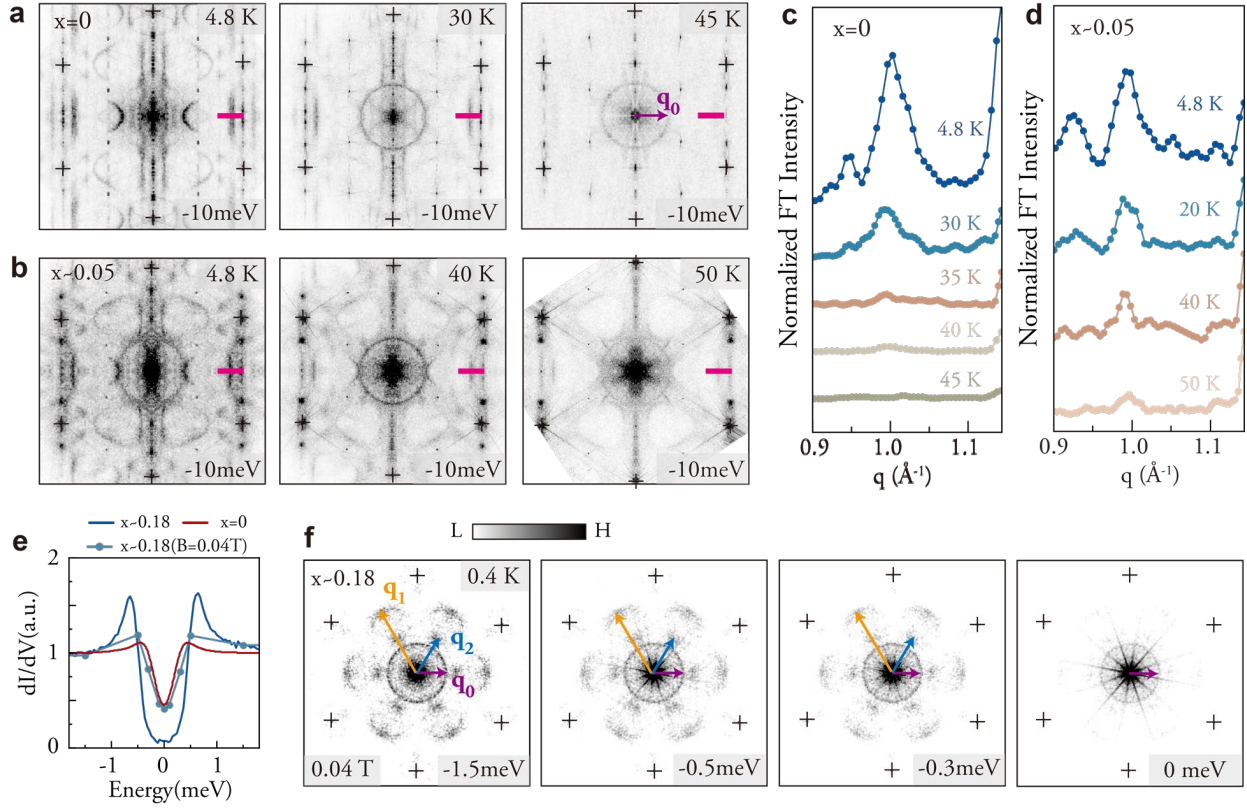

**Figure 4. Temperature-evolution of the unidirectional quasiparticle interference and momentum-dependence of Cooper pairing. a,** Temperature-dependent $L(q,-10\text{meV})$ maps for x = 0 samples. While the Sb-derived interference pattern ($q_0$) persists to high temperatures, those related to the vanadium bands ($q_1$ to $q_3$) fade at approximately 40 K. **b,** Temperature-dependent $L(q,-10\text{meV})$ maps for x ≈ 0.05 samples, where signals of $q_1$ and $q_3$ survive to higher temperatures. **c, d,** FT line-cuts taken along the pink lines in **a** and **b**. The data are normalized to the amplitude of the associated Bragg peaks. **e,** Superconducting spectra for x = 0 and x ≈ 0.18 at 0.4 K. **f,** Measured Bogoliubov quasiparticle interference for x ≈ 0.18 at 0.04 T, showing more remaining spectral weights on the Sb-$p$ band deeply inside the gap. A quantitative analysis of the energy-dependent spectral weight on the V-bands and Sb band is presented in Extended Data Fig. 9, for 0T, 0.04 T and 2 T. STM setup condition: **a** left: $V_s$ = -10 mV, $I_t$ = 5 nA, $V_m$= 3 mV; middle: $V_s$ = -10 mV, $I_t$ = 3 nA, $V_m$= 3 mV; right: $V_s$ = -10 mV, $I_t$ = 3 nA, $V_m$= 3 mV; **b** left: $V_s$ = -10 mV, $I_t$ = 4 nA, $V_m$= 2 mV; middle: $V_s$ = -10 mV, $I_t$ = 4 nA, $V_m$= 3 mV; right: $V_s$ = -10 mV, $I_t$ = 3 nA, $V_m$= 3 mV; **e**: x=0: $V_s$ = -2 mV, $I_t$ = 250 pA, $V_m$ = 0.2 mV, x≈0.18: $V_s$ = -2 mV, $I_t$ = 200 pA, $V_m$ = 0.1 mV, **f**: $V_s$ =-8 mV, $I_t$=400 pA, $V_m$ = 0.2 mV, T = 0.4 K, B = 0.04 T.

## Method

**Single crystal growth and characterization.** Single crystals of CsV$_{3-x}$Ti$_x$Sb$_5$ were synthesized via a self-flux growth method similar to the previous reports (19). The as-grown CsV$_{3-x}$Ti$_x$Sb$_5$ single crystals are stable in the air. The chemical composition of the crystals was determined by inductively coupled plasma-atomic emission spectroscopy (ICP-AES) and STM topographic images. Electrical transport measurements were carried out in a Quantum Design physical property measurement system (PPMS-14T). Electrical transport measurement in a conventional 4-lead configuration was realized by attaching four-platinum wires to the (001) surface of the sample. Contacts were made by DuPont 4929N conductor paste.

**STM measurements**: The STM data above 1 K were acquired using a commercial CreaTec low temperature STM system, and the sub-kelvin studies were performed on a Unisoku USM1300 system. The single crystals were cleaved *in situ* in cryogenic ultrahigh vacuum at $T \sim 30$ K and immediately inserted into the STM head. PtIr tips were used and their qualities were tested on the surface of single crystalline Au (111) prior to measurements. Spectroscopic data were acquired by the standard lock-in technique at a frequency of 987.5 Hz, under modulation voltage of 2-5 mV.

The QPI measurements presented in the main text were performed in single $C_2$ domains. The domain size is about tens to hundreds nanometer in $x = 0$ and $x \approx 0.05$ samples, and becomes smaller (~10- 20 nm) for $x \approx 0.12$ samples (Supplementary Note 2). In this work, we measured three $x = 0$, four $x \approx 0.05$, two $x \approx 0.12$ samples (from one large batch) and four $x \approx 0.18$ samples. The normal-state dispersion has been checked on at least two samples for $x = 0$, $x \approx 0.05$ and $x \approx 0.18$, and the BQPI results have been checked on two samples for $x = 0$ and $x \approx 0.18$. Before performing SI-STM measurements, we always checked the local Ti concentration in the field of view (FOV) with topographic images taken at 1.5 V (or -1.0 V), as listed in Supplementary Table 1.

For a better representation of the low-$q$ scattering events of the vanadium bands, we included as few Sb-site defects as possible in the FOVs, which otherwise would produce circular-shaped interference patterns (of Sb-$p$ orbital) and random defect-phase information at small $q$ in the FTs. The Lawler–Fujita drift-correction algorithm (59) was used for spectroscopic mapping to remove the drift effects. The Fourier transforms, $L(\boldsymbol{q}, eV)$, have been rotated to align the lattice vector in the same way for each dataset, and then cropped to highlight the main features in the first Brillouin zone. The full-scale maps are shown in Supplementary movies.

In STM measurements, the tunneling current can be generally written as
$$I(\boldsymbol{r}, V_0) = \frac{4\pi e}{\hbar} e^{-\kappa z(\boldsymbol{r})} \int_0^{eV_0} n(\boldsymbol{r}, \omega) d\omega \quad \text{(M1)}$$

Here $V_0$ is the tip-sample bias voltage, $z(\mathbf{r})$ is the tip-surface distance, $\kappa$ is related to the work functions of both the tip and the sample, and $n(\mathbf{r}, \omega)$ is the desired electron density of states of the sample. In the constant current mode, by fixing a certain current $I_0$ at a given bias voltage $V_0$, the tip-surface distance $z(\mathbf{r})$ should be location-dependent if $\int_0^{eV_0} n(\mathbf{r}, \omega) d\omega$ is not a spatial uniform term, which in turn will impede an accurate determination of $n(\mathbf{r}, \omega)$. This systematic error is known as the 'set-point' effect, and then, the measured $g(\mathbf{r}, E = eV)$ is given by

$$g(\mathbf{r}, eV) = \frac{eI_0}{\int_0^{eV_0} n(r,\omega)d\omega} n(\mathbf{r}, eV) \tag{M2}$$

It is clear that the quantity $g(\mathbf{r}, eV)$ depends on the choice of $V_0$. We can normalize the measured $g(\mathbf{r}, eV)$ by using the Feenstra function (39):

$$L(\mathbf{r}, eV) = \frac{g(\mathbf{r},eV)}{I/V} = \frac{eV}{\int_0^{eV} n(r,\omega)d\omega} n(\mathbf{r}, eV) \tag{M3}$$

One can see that the integral in the denominator still exists; however, $L(\mathbf{r}, eV)$ is not a function of $V_0$. This treatment thus allows consistent comparison between results taken on different samples and with different set-points.

In general, the QPI data detect electronic states that have negligible group velocity in the z-axis (usually the $k_z = 0$ or $k_z = \pi/c$ slice in a 3D case; 60, 61). For $x = 0$ and $x \approx 0.05$, we find good agreement between the experimental data and band calculation with $2 \times 2 \times 2$ CDW at both $k_z = 0$ and $k_z = \pi/2c$ (here $k_z = \pi/2c$ is the BZ boundary due to the doubling of the unit-cell along the z direction). For $x \approx 0.18$, we find that the QPI dispersion is more consistent with the calculation at $k_z = \pi/c$ (Supplementary Note 5). We do not find scattering channels that arise from the in-plane CDW folding in the QPI data, probably because the in-plane folding of vanadium is too weak (41) to generate visible scattering signals (Supplementary Note 6). The energy dispersion of the Sb-*p* band is presented in Supplementary Note 7, where kink feature is observed at ~ 30meV. However, the kink on the Sb-*p* band is rather isotropic since the related QPI pattern is always circular-shaped without visible angular-dependence (Fig. 3c and Supplementary Note 4).

**BQPI analysis**: Owing to the small superconducting energy gap (~0.5 meV), the limited energy-resolution at 0.4 K (~0.16 meV) and the discrete nature of QPI patterns in the normal state, it is not easy to precisely determine the superconducting gap function $\Delta(\mathbf{k})$. However, quantitative analysis is still informative to clarify the superconducting gap on the vanadium bands. To do this, we first suppress the central core in the FT with Gaussian backgrounds, and then integrate the total spectral weight in the relevant q-space ranges around $q_0$, $q_1$ and $q_2$ (Extended Fig. 9) and obtain the averaged spectral weight per pixel ($\bar{g}(\mathbf{q}_i, eV)$) as a function of energy. To illustrate the difference between $\bar{g}(\mathbf{q}_i, eV)$ on the vanadium and Sb bands, we calculate a ratio of $\Delta\bar{g}(eV) =$ $\frac{[\bar{g}(q_1,eV)+\bar{g}(q_2,eV)]/2-\bar{g}(q_0,eV)}{[\bar{g}(q_1,eV)+\bar{g}(q_2,eV)]/2+\bar{g}(q_0,eV)}$, where $\bar{g}(q_1,eV)$ and $\bar{g}(q_2,eV)$ are the averaged spectral weight per pixel on the V bands and $\bar{g}(q_0,eV)$ is that on the Sb band.

When the superconductivity is fully suppressed at 2T, $\Delta\bar{g}(eV)$ is expected to show no change in the energy window of the superconducting gap. This is exactly what we find for the case of 2T. At 0 T and 0.04 T, the spectral weight near the Fermi energy will be suppressed due to the opening of the superconducting gap. We find that $\Delta\bar{g}(eV)$ shows a clear drop to negative values inside the superconducting gap, suggesting a faster suppression of spectral weight on the V bands inside the gap. This is further supported by a direct comparison in FT linecuts across $q_0$, $q_1$ and $q_3$. These observations indicate that the superconducting gap magnitude on the V bands, at least, is similar to that on the Sb band. For the measurement at 0.04 T, the vortex region is small in the FOV and does not affect the BQPI data, which has been checked by area-restricted Fourier analyses (Supplementary Note 8).

**Band calculations:** First-principles calculations were performed by using the projected augmented-wave method implemented in the Vienna ab initio simulation package. The generalized gradient approximation (GGA) of the Perdew-Burke-Ernzerhof type was utilized to treat the exchange-correlation interaction with a plane-wave kinetic energy cut-off of 350 eV. The Hellmann-Feynman force tolerance criterion for convergence was 0.01 eV/Å. We used zero damping DFT-D3 method to describe the van der Waals interaction between adjacent layers. The Brillouin zone was sampled with a 16×16×9 Monkhorst-Pack grid for structural optimization and electronic structure calculation. For the normal unit cell, the lattice constants are 5.451 Å (a) and 9.228 Å (c) after full relaxation. In the 2×2×1 supercell, the tri-hex or SoD distortion of V kagome layers was initialized and then we performed the relaxation of atomic positions. For the 2×2×2 supercell, we used the experimental lattice parameters determined by XRD at 18 K from Ref. 41. The 5×5×5 and 3×3×3 Monkhorst-Pack grids were used to sample the Brillouin zone of 2×2×1 and 2×2×2 supercells, respectively. The Bloch states determined from the DFT calculations of bulk were projected to the V-3$d$ and Sb-5$p$ orbitals to obtain Wannier representations as implemented in Wannier90 (62). On the basis of the tight-binding model established by maximally localized Wannier function (MLWF), the surface states are calculated by the iterative Green's function as implemented in WannierTools package (63). Note that counting Ti atoms in topographic images gives a local Ti concentration of 0.15 (Supplementary table S1) for the $x \approx 0.18$ case, thus we constructed a configuration of a 4×4×1 supercell reconstruction with two-Ti-atom substitution (with $x \approx 0.13$), where the fixed cell volume and 4×4×5 $k$-meshes were used for the structure relaxation. The calculated bands in the supercell were unfolded into the Brillouin zone of the corresponding primitive cell within post-processing VASPKIT package (64). The phonon dispersion was calculated by finite displacement method as implemented in the PHONOPY code (65). The DOS of phonons in the pristine phases was calculated with a 3×3×2 supercell, while the 2×2×1 CDW phases were calculated with a 2×2×2 supercell.

## Acknowledgements

We thank Rafael M Fernandes, Vidya Madhavan, Kun Jiang, Xianxin Wu, Zengyi Du, Rahul Sharma, Junfeng He and Yiming Sun for valuable discussions. We thank Yingying Peng for providing XRD data for the band structure calculation. This work is supported by the National Key R&D Program of the MOST of China (Grants Nos. 2022YFA1602600 and 2017YFA0303000), the National Natural Science Foundation of China (Grants Nos. 11888101, 12034004, 12074364 and 52261135638), the Strategic Priority Research Program of Chinese Academy of Sciences (Grant No. XDB25000000), the Anhui Initiative in Quantum Information Technologies (Grant No. AHY160000), and the Collaborative Innovation Program of Hefei Science Center, CAS (Grant No. 2019HSC-CIP007). Z.Y.W. is also supported by the Fundamental Research Funds for the Central Universities (WK3510000012 and WK3510000011).

## Author contributions

Z.Y.W. and X.H.C. conceived the experiments and supervised the project. P. W., Y. T., Z. W., S. Y. and Y. Z. performed STM experiments and data analysis with assistance of W. M., Z. Liang, X. Z. and L. S. H. L., J. Y. and Z. X. synthesized and characterized the samples. Z. Li, W. M., Y. Y. and Z. Q. performed band calculations. Z.Y.W., Z. X., T.W. and X.H.C. interpreted the results and wrote the manuscript. All authors discussed the results and commented on the manuscript.

## Competing financial interests

The authors declare no competing interests.

## Data availability

The data supporting the findings of this study are available from the corresponding author upon reasonable request. Source data will be provided with this paper.

## Code availability

The code used for STM data analysis is available from the corresponding author upon reasonable request.

**Extended Data Figure 1**

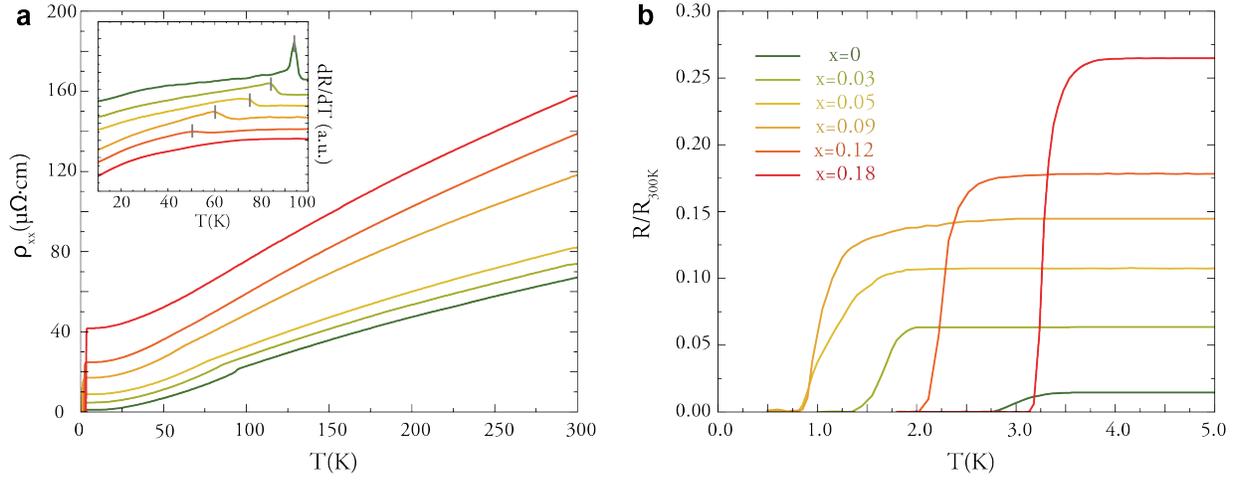

**Extended Data Figure 1. Transport characterization of the Ti-substituted CsV$_3$Sb$_5$ samples. a,** Temperature-dependent in-plane resistivity of CsV$_{3-x}$Ti$_x$Sb$_5$ single crystals, with x=0, 0.03, 0.05, 0.09, 0.12 and 0.18. Inset: temperature-derivative of the in-plane resistivity showing the charge density wave transition. **b,** Zoomed-in view of the low-temperature resistivity showing the superconducting transition $T_c$. In this study, $T_c$ is defined as the zero-resistance temperature.

**Extended Data Figure 2**

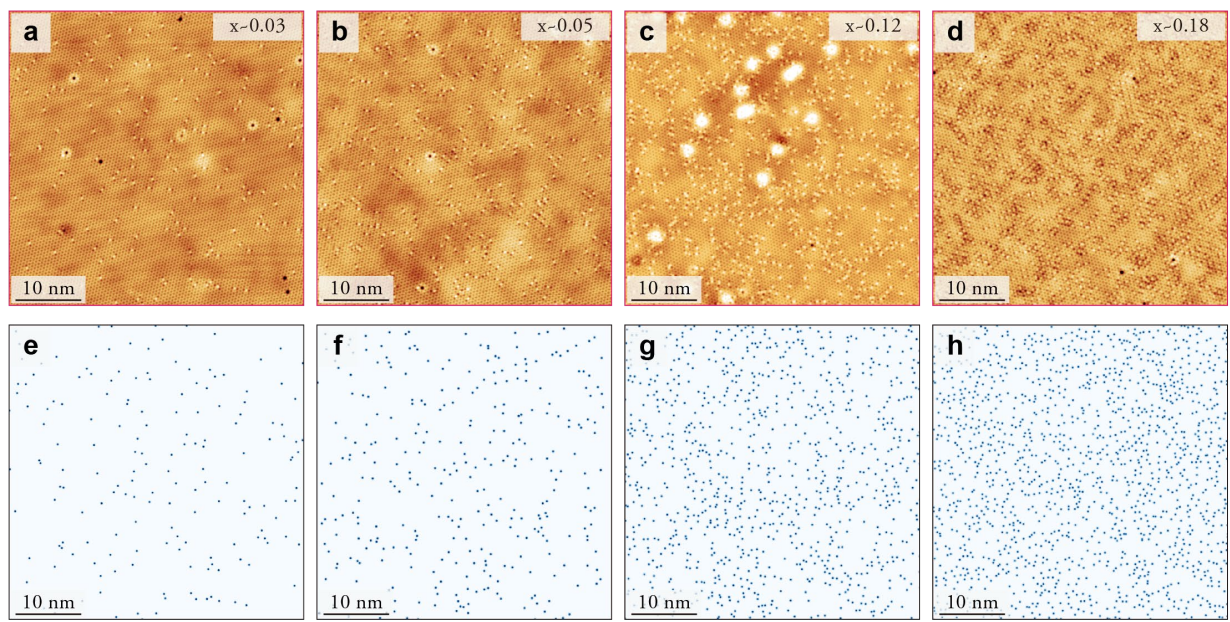

**Extended Data Figure 2. Determination of the local Ti concentration. a** to **d,** Typical topographic images of CsV$_{3-x}$Ti$_x$Sb$_5$. The bright protrusions represent Ti atoms that replace V atoms in the underneath kagome layer. **e** to **h,** Distributions of Ti atoms in the field of view shown in **a** to **d**. The numbers of Ti atoms by counting are 172, 312, 913, and 1410, respectively. Set-point conditions: **a**: $V_s$ = 1.5 V, $I_t$ = 6 nA; **b**: $V_s$ = 1.5 V, $I_t$ = 4 nA; **c**: $V_s$ = 1.5 V, $I_t$ = 2 nA; **d**: $V_s$ = 1.5 V, $I_t$ = 5 nA.

**Extended Data Figure 3**

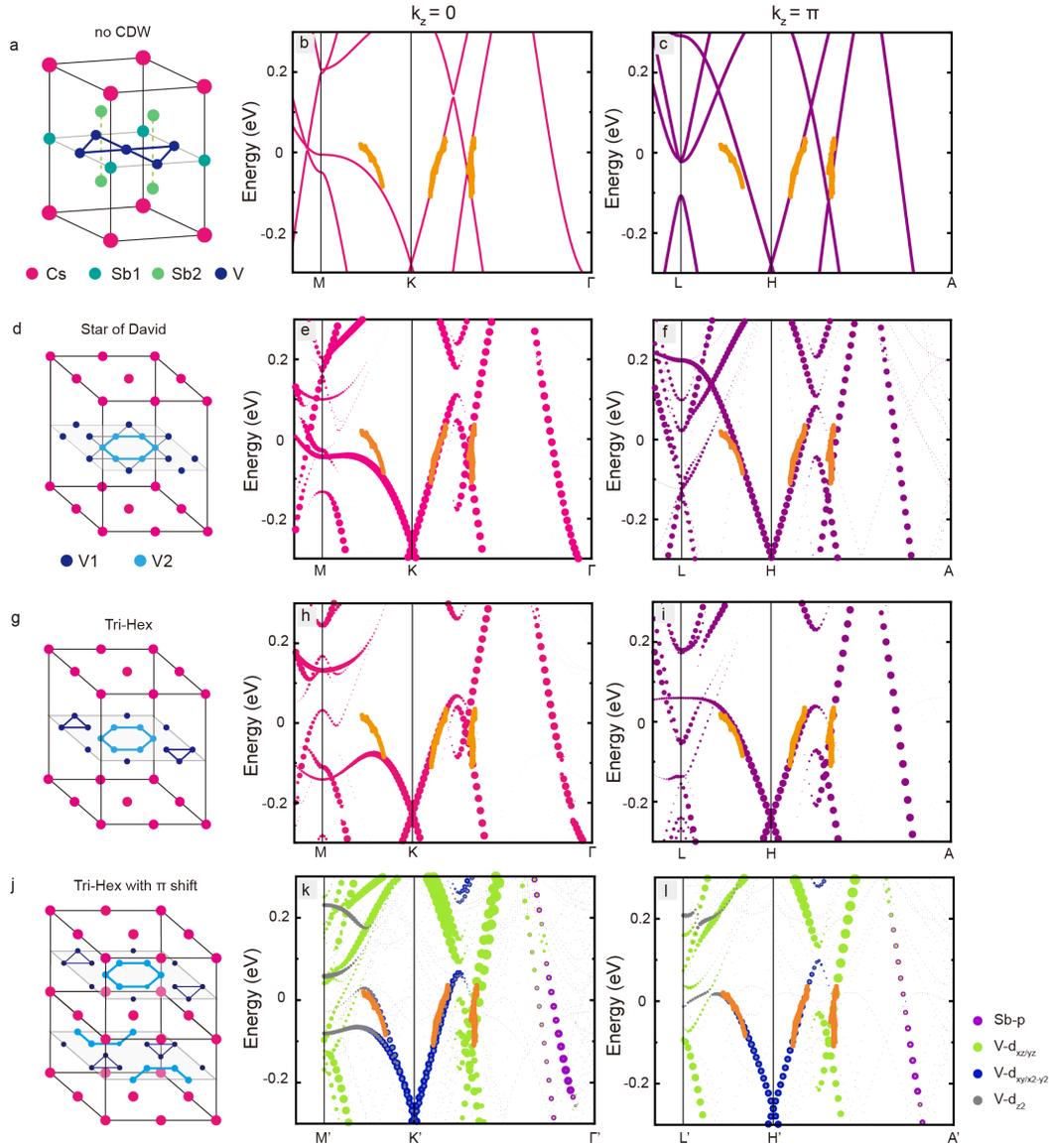

**Extended Data Figure 3. Comparison of the measured $k_i(\omega)$ to the calculated band dispersion using different CDW configurations.** The orange dots denote the measured QPI dispersion for x = 0 and x ≈ 0.05. **b-c**, Calculated band structure with the pristine crystal structure in **a** at $k_z = 0$ and $k_z = \pi/c$. **e-f**, Calculated band structure with SoD reconstruction in **d** at $k_z = 0$ and $k_z = \pi/c$. **h-i**, Calculated band structure with Tri-Hex reconstruction in **g** at $k_z = 0$ and $k_z = \pi/c$. **j-l**, Calculated band structure with 2 ×2 ×2 CDW reconstruction in **g** (Tri-Hex with π-shift) at $k_z = 0$ and $k_z = \pi/2c$ (here $\pi/2c$ is the boundary of the extended cell). *Γ'*, *K'*, *M'*, *A'*, *H'*, *L'* are the high-symmetry points of the folded BZ due to 2 ×2 ×2 CDW reconstruction which corresponds to *Γ*, K, M, A, H, L in the unfolded BZ.

**Extended Data Figure 4**

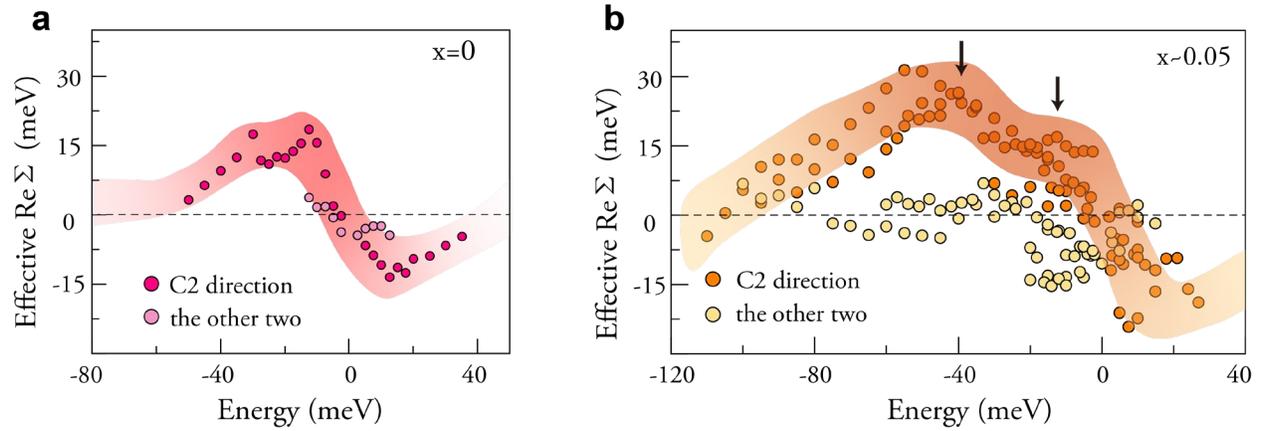

**Extended Data Figure 4. Effective real part of self-energy for the $k_1^\delta(\omega)$ band along different lattice directions.** Effective real part of quasiparticle self-energy, $\Sigma'(\mathbf{k},\omega)$, extracted from the measured dispersion $k_1^\delta(\omega)$ for x = 0 (**a**) and x ≈ 0.05 (**b**) samples. Band dispersion obtained by calculations is used as the 'bare' band and slightly shifted in energy to match the measured $k_F$. Peaks located at approximately 15 and 30-40 meV are observed along the $C_2$ lattice direction, while there is no obvious feature along the other two lattice directions. Thick diffuse lines in **a** and **b** are visual guides. For $x = 0$ samples, interference signals along the off-$C_2$ direction are weak at $|E| > 20$ meV.

**Extended Data Figure 5**

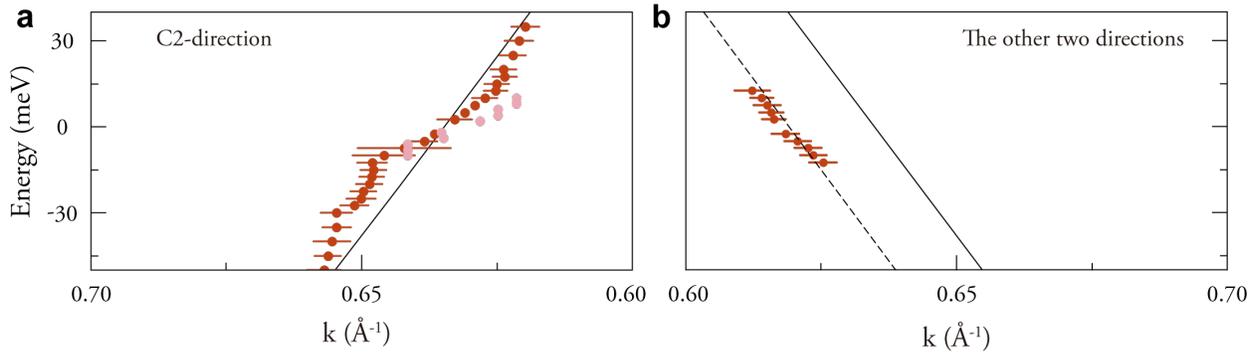

**Extended Data Figure 5. Measured $k_1^\delta(\omega)$ dispersion for x=0 samples.** Kinks at approximately 15 and 30 meV are observed only along the $C_2$-direction (**a**); along the remaining two lattice directions, the measured $k_1^\delta(\omega)$ roughly follows the calculated 'bare' band without additional features (**b**). The determination of $k_1^\delta(\omega)$ by fitting is shown in Supplementary Note 3. In the x = 0 samples, a small Fermi surface deformation of the δ band is observed, that is, $k_1^\delta(E_F)$ is different along the $C_2$-direction than along the remaining two lattice directions (dashed line in **b**), indicating a possible *q=0* Pomeranchuk instability (25, 26). However, this effect is very weak in the x ≈ 0.05 samples (Fig. 3e).

**Extended Data Figure 6**

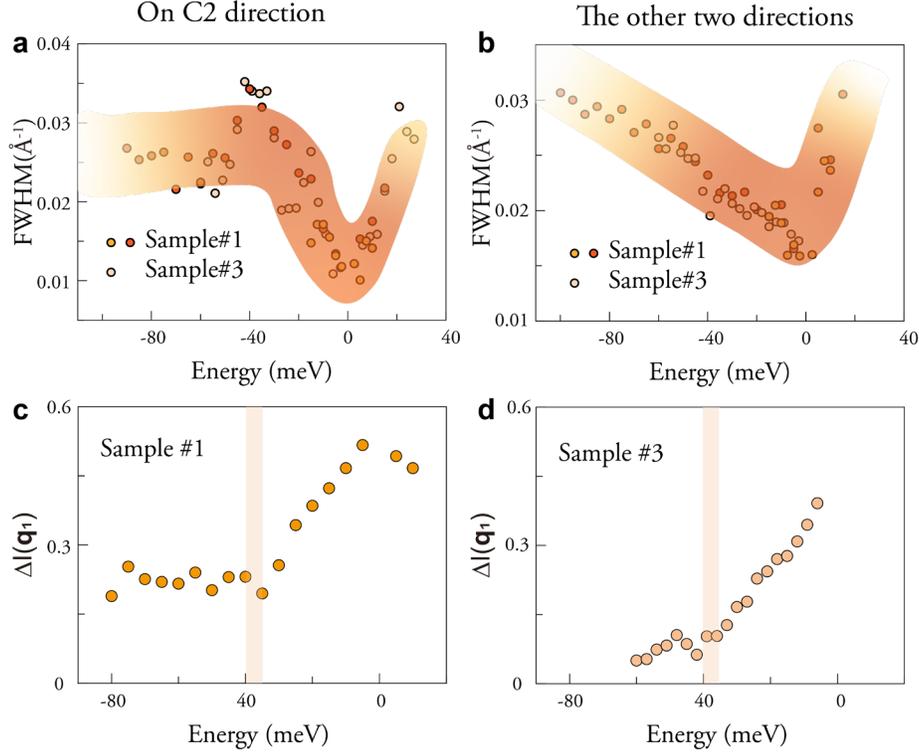

**Extended Data Figure 6. Linewidth and peak height in the STM momentum distribution curves for x ≈ 0.05.**
**a, b,** Phenomenological full-width-half-maximum ($\Delta k = \Delta q/2$) of $L(q_1, eV)$ extracted by Lorentzian fits along the $C_2$ direction (a) and off-$C_2$ direction (b). The fitting details are shown in SI Note 3. In principle, the peak widths can be related to the quasiparticle lifetime and thus the imaginary part of the self-energy. A remarkable change in the peak width can be found along the $C_2$ direction at -40 meV, which corresponds to the kink position in the dispersion. This is consistent with $C_2$-symmetric band renormalization. **c, d,** The anisotropy of the FT peak height at $q_1$ along the $C_2$ direction and off-$C_2$ direction, defined as $\Delta I(q_1) = \frac{I(q_1^{C2}) - I(q_1^{other})}{I(q_1^{C2}) + I(q_1^{other})}$, for two different samples. The anisotropy of the peak height shows a significant increase at $|E| < 40$ meV, again consistent with $C_2$-symmetric band renormalization by electron-mode coupling.

**Extended Data Figure 7**

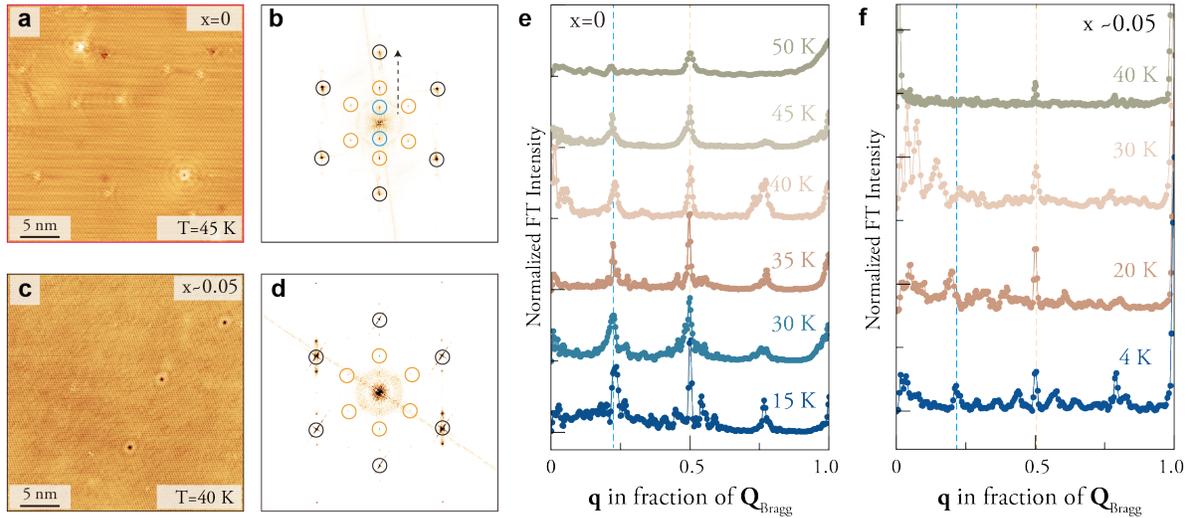

**Extended Data Figure 7. Temperature dependence of the 1$Q$-4a$_0$ charge order. a,** Atomically resolved STM topography of the Sb-terminated surface at 45 K, for x = 0. **b,** The FT of the topography shown in **a**. **c, d,** STM topography obtained at 45 K for x ≈ 0.05 samples and its FT. The dark circles, orange circles and blue circles denote the Bragg peaks, triple-$Q$ 2a$_0$ × 2a$_0$ CDW peaks and 1$Q$-4a$_0$ charge order peaks, respectively. **e, f**, Temperature-dependent FT linecuts of STM topographic images along the 1$Q$-4a$_0$ ordering direction. All the linecuts are normalized with the amplitude of the associated Bragg peaks for comparison. One can find that the onset temperature of the 1$Q$-4a$_0$ order is suppressed in the x ≈ 0.05 samples.

**Extended Data Figure 8**

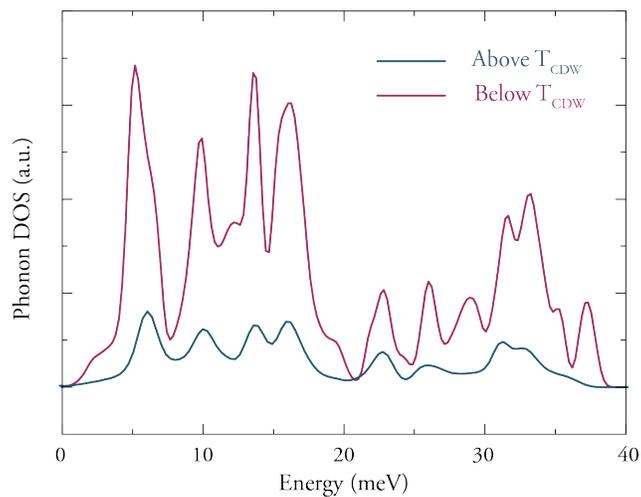

**Extended Data Figure 8. Calculated phonon density of states.** The phonon density of states shows a series of peaks in similar energy windows where we see kinks in the electron dispersion: some near 15 meV and others between 30 and 40 meV. The amplitudes of these peaks are largely enhanced when entering the CDW phase (calculated for Tri-Hex configuration; see Method).

**Extended Data Figure 9**

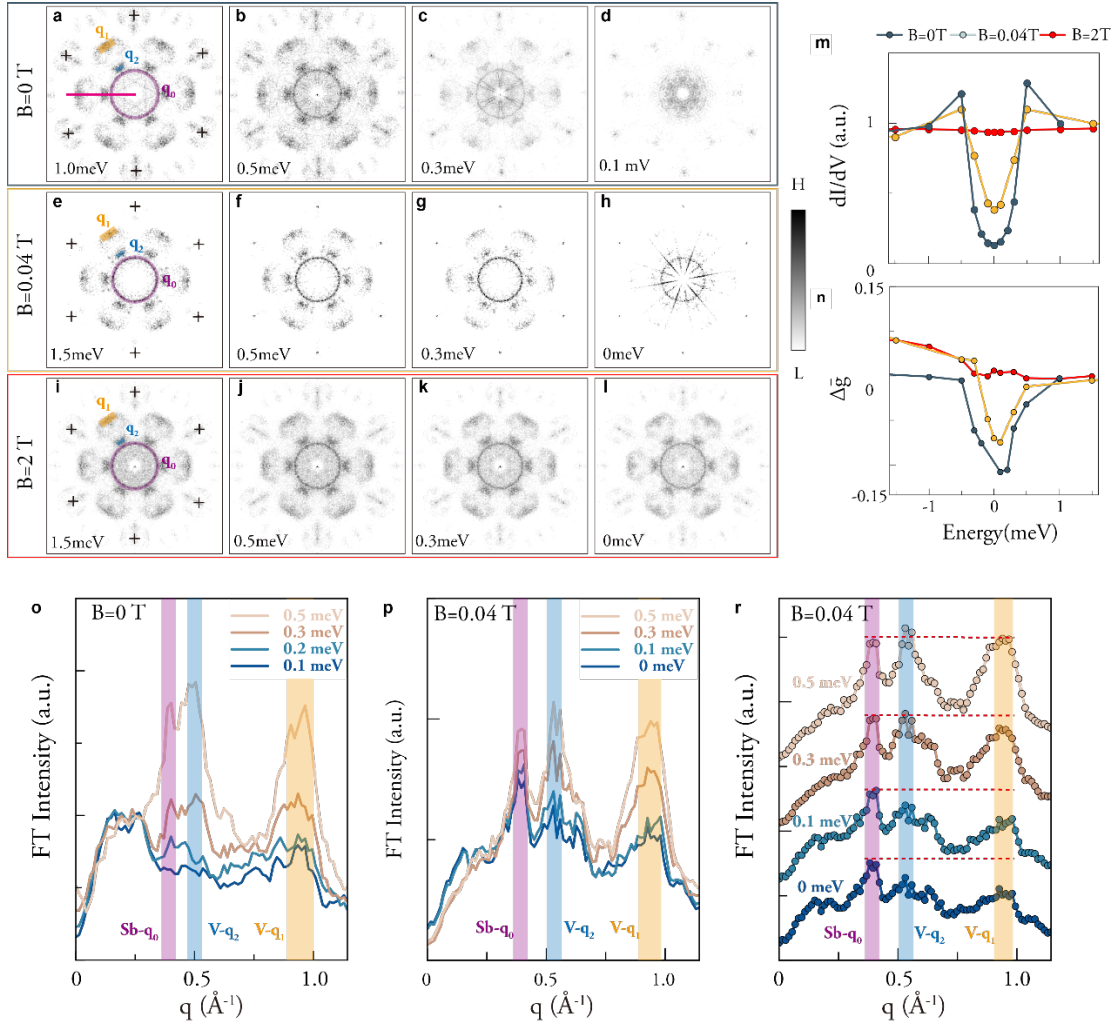

**Extended Data Figure 9. QPI maps at 0T, 0.04T and 2T for x ≈ 0.18. a-d,** Measured $g(q, eV)$ at 0.4 K with zero magnetic field. **e-h,** Measured $g(q, eV)$ at 0.4 K with 0.04 T perpendicular magnetic field. **i-l,** Measured $g(q, eV)$ at 0.4K with 2T perpendicular magnetic field (above $H_{C2}$). Raw data were symmetrized to increase the signal-to-noise ratio, and the central core is suppressed using a Gaussian. The $q$-space range for spectral weight analysis is color-marked in **a**, **e**, and **i**. **m**, averaged $dI/dV$ spectra at 0T, 0.04 T and 2 T. **n**, Difference between the spectral weight $\bar{g}(q_i, eV)$ on the vanadium and Sb bands as a function of energy, $\Delta\bar{g}(eV) = \frac{[\bar{g}(q_1,eV)+\bar{g}(q_2,eV)]/2-\bar{g}(q_0,eV)}{[\bar{g}(q_1,eV)+\bar{g}(q_2,eV)]/2+\bar{g}(q_0,eV)}$, where $\bar{g}(q_1, eV)$ and $\bar{g}(q_2, eV)$ are the averaged spectral weight per pixel on the V bands and $\bar{g}(q_0, eV)$ is that on the Sb band (see Method). The drop to negative values of $\Delta\bar{g}(eV)$ inside the superconducting gap at 0T and 0.04T suggests a faster suppression of spectral weight on the V bands in the superconducting state. **o**, **p**, FT line-cuts along the pink line in **a**, showing the evolution of $q_0$, $q_1$ and $q_3$ with energy for B=0T and 0.04T, respectively. **r,** Same line-cuts as shown in p but vertically shifted to highlight the change of $q_0$, $q_1$ and $q_3$ within the superconducting gap. It is clear that the peak intensities of $q_1$ and $q_3$ (V-bands) drop faster than that of $q_0$ (Sb-band).

**Extended Data Figure 10**

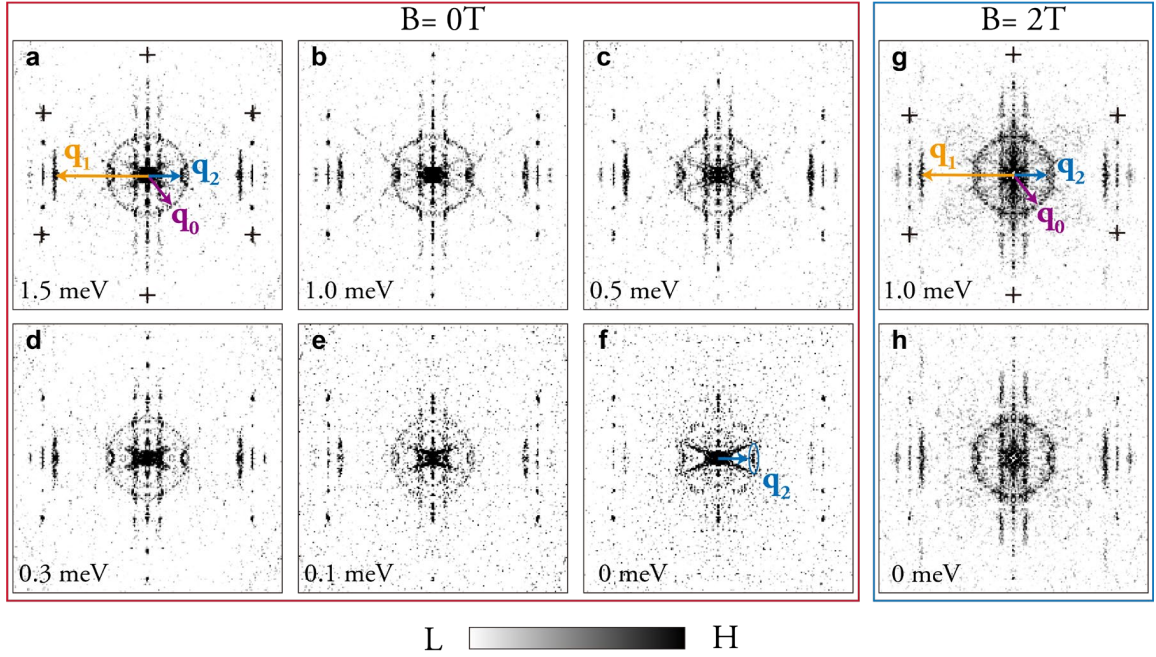

**Extended Data Figure 10. QPI maps at 0T and 2T for x = 0. a-f,** Measured $g\,(q, eV)$ at 0.4 K with zero magnetic field. The spectra weight located at the vector of $q_2$ persists to zero energy inside the superconducting gap, suggesting a gap minimum on the V- $d_{xy}/d_{x^2-y^2}$ band near the $M$ point. **g-h,** Measured $g\,(q, eV)$ at 0.4 K with perpendicular magnetic field of 2 T (above $H_{C2}$).